\documentclass[12pt]{iopart}
\usepackage{amsfonts}
\usepackage{amsbsy}
\usepackage{amssymb}
\usepackage{amsthm}
\usepackage{amsgen}
\usepackage{epsfig}

\theoremstyle{plain}
\newtheorem{Th}{Theorem}
\newtheorem{Def}{Definition}
\def\eqref#1{(\ref{#1})}
\theoremstyle{remark}
\newtheorem{Rem}{Remark}
\newcommand{\PP}{{\mathbb P}}
\newcommand{\RR}{{\mathbb R}}
\newcommand{\EE}{{\mathbb E}}
\newcommand{\ZZ}{{\mathbb Z}}

\newcommand{\calC}{{\mathcal C}}

\newcommand{\calA}{{\mathcal A}}

\newcommand{\cQ}{{\mathcal Q}}

\newcommand{\bx}{\bi{x}}
\newcommand{\bX}{\bi{X}}
\newcommand{\boe}{\bi{e}}
\newcommand{\bN}{\bi{N}}
\newcommand{\bn}{\bi{n}}
\newcommand{\by}{\bi{y}}
\newcommand{\bu}{\bi{u}}
\newcommand{\bv}{\bi{v}}
\newcommand{\bnu}{\boldsymbol{\nu}}
\newcommand{\bW}{\bi{W}}
\newcommand{\bZ}{\bi{Z}}

\newcommand{\gp}{{\mathfrak p}}
\newcommand{\gq}{{\mathfrak q}}

\begin{document}

\article[Asymptotic lattices and their integrable reductions I]{}{Asymptotic 
lattices and their integrable reductions I:
the Bianchi and the Fubini-Ragazzi lattices}

\author{A Doliwa\dag, M Nieszporski\ddag, P M Santini\S}

\address{\dag\ Instytut Fizyki Teoretycznej, Uniwersytet Warszawski, 
ul. Ho{\.z}a 69, 00-681 Warszawa, Poland}

\address{\ddag\ Instytut Fizyki Teoretycznej, Uniwersytet w Bia{\l}ymstoku,
ul. Lipowa 41, 15-424 Bia{\l}ystok, Poland}

\address{\S Dipartimento di Fisica, Universit\`a di Roma ,,La Sapienza'', 
Istituto Nazionale di Fisica Nucleare, Sezione di Roma,
P-le Aldo Moro 2, I--00185 Roma, Italy}

\eads{\mailto{doliwa@fuw.edu.pl}, \mailto{maciejun@fuw.edu.pl},
\mailto{maciejun@uwb.edu.pl}, \mailto{paolo.santini@roma1.infn.it}}

\begin{abstract}
We review recent results on asymptotic lattices and their integrable
reductions. We present the theory of general asymptotic lattices in $\RR^3$
together with the corresponding theory of their Darboux-type
transformations. Then we study the discrete analogues of the Bianchi 
surfaces and their transformations. Finally, we present the corresponding
theory of the discrete analogues of the isothermally-asymptotic 
(Fubini--Ragazzi) nets. 
\end{abstract}

\maketitle
\section{Introduction}
One of the best known example of integrable geometries is provided by
asymptotic nets  on surfaces of constant
curvature, which are described by the sine-Gordon equation \cite{Bianchi}. 
It turns out that asymptotic nets 
on surfaces in $\EE^3$ provide other classes of integrable geometries, 
for example, Bianchi surfaces \cite{Bianchi,Levi},
affine spheres \cite{Tzitzica}, isothermally-asymptotic nets (Fubini-Ragazzi nets) 
\cite{Fubini,Ragazzi,NieszporskiA}.

The discrete analogues of asymptotic nets (asymptotic lattices) have 
been proposed long time ago 
by Sauer \cite{Sauer}.
He also considered the "discrete pseudospherical surfaces" whose study 
was recently undertaken by
Bobenko and Pinkal \cite{BoPi-DP} from the point of view of integrable 
systems. Recently Bobenko and Schief introduced 
the  discrete  analogue of (indefinite) affine spheres 
\cite{BoSchA,BoSchI}, which is described by the discrete
analogue of the Tzitzeica equation. 

The integrability aspects of generic asymptotic lattices were the subject 
of studies of Nieszporski
\cite{NieszporskiA} and Doliwa \cite{DoliwaA}. In particular, in the paper 
of Doliwa, the theory of asymptotic
lattices and their transformations was considered as a part of the theory 
of quadrilateral lattices (the discrete
analogues of conjugate nets); for information about the quadrilateral 
lattices, their transformations and
reductions see \cite{DoSaMQL,DoSaMa,DoManSa,KoSchi,DoSaS}.
More recently Nieszporski, Doliwa and Santini introduced
the integrable discrete analogue of the Bianchi surfaces 
\cite{NieszporskiA,NDS}. Also the integrable
discrete analogue of the isothermally-asymptotic nets 
(Fubini-Ragazzi nets), which includes the discrete
affine spheres as a particular integrable subreduction, 
was introduced by Nieszporski in \cite{NieszporskiA,NieszporskiD}.

In this paper we present the theory of asymptotic 
lattices and their integrable reductions from a unified perspective. 
In addition to the results already known in the literature,
we develop the theory of transformations of the discrete Bianchi surfaces 
and the  discrete Fubini-Ragazzi nets. 

The paper is organized as follows. Section \ref{sec:A} is devoted to general
theory of asymptotic lattices and their transformations. In Section
\ref{sec:R} we present the discrete analogues of the Bianchi and
Fubini--Ragazzi reductions of the asymptotic nets. In 
\ref{app:QL} we collected some results from the theory of quadrilateral
lattices which are used in this review. In \ref{app:lg} we present basic
notions of the line geometry of Pl\"ucker.

We use the following notation: given a function $f$ defined on the
two dimensional integer lattice $\ZZ^2\ni(m_1,m_2)$, we denote by
$f_{(\pm i)}$, $i=1,2$, the function $f$ of the shifted arguments, i.e.,  
$f_{(\pm 1)}(m_1,m_2)=f(m_1\pm 1,m_2)$,
$f_{(\pm 2)}(m_1,m_2)=f(m_1,m_2\pm 1)$ and $f_{(12)}(m_1,m_2)=f(m_1+1,m_2+ 1)$. 
We make use also the following
difference operators $\Delta_i f=f_{(i)}-f$.

\section{Asymptotic lattices and W--congruences}
\label{sec:A}
In this section we present the theory of general asymptotic lattices. 
For these lattices, characterized by linear difference equations (equations
\eqref{eq:da1} or \eqref{DME1N} below), there exist Darboux-type
transformations whose superposition satisies the permutability property.
Therefore they can be coined {\bf integrable lattices}.

\subsection{Asymptotic lattices}
\begin{figure}
\begin{center}
\epsfbox{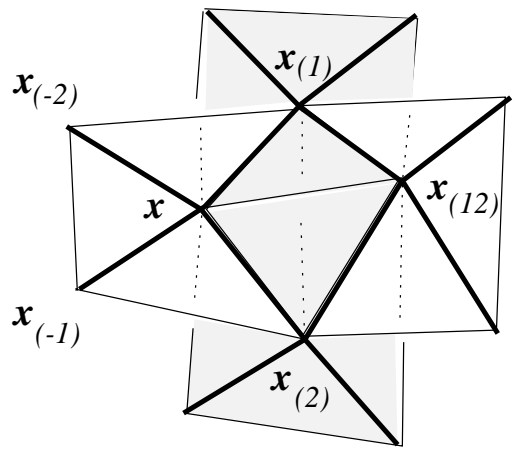}
\end{center}
\caption{Asymptotic lattice}
\label{fig:d-as2}
\end{figure}

The asymptotic lattice is defined like in the continuous case
and, roughly speaking, it is a two
dimensional lattice such that osculating
planes of the parametric curves coincide in the intersection point
(see Figure \ref{fig:d-as2}).
\begin{Def}[\cite{Sauer}] \label{def:d-as}
An {\it asymptotic lattice (or discrete asymptotic net)} is a mapping 
$\bx:\ZZ^2\to \RR^3$
such that any point $\bx$ of the lattice is coplanar with its four nearest 
neighbours $\bx_{(1)}$, $\bx_{(2)}$, $\bx_{(-1)}$ and $\bx_{(-2)}$.
\end{Def}
\begin{Rem}
The common plane of the five points  $\bx$, $\bx_{(1)}$, $\bx_{(2)}$, 
$\bx_{(-1)}$ and $\bx_{(-2)}$ is the tangent plane of the lattice at $\bx$.
\end{Rem}
\begin{Rem}
Throught the paper we consider only non-degenerate asymptotic lattices,
i.e., for every $\bx$ the three vectors $\bx_{(1)}-\bx$, $\bx_{(2)}-\bx$ and
$\bx_{(12)}-\bx$ are linearly independent.
\end{Rem}

Algebraically, the asymptotic lattice condition can be rewritten in 
the form of the following linear system \cite{BoSchI,NieszporskiA}
\begin{equation} \label{eq:da1}
\eqalign{
\bx_{(11)}-\bx_{(1)} &= A(\bx_{(1)}-\bx) + P(\bx_{(12)}-\bx_{(1)}),\\
\bx_{(22)}-\bx_{(2)} &= B(\bx_{(2)}-\bx) + Q(\bx_{(12)}-\bx_{(2)}),
}
\end{equation}
which gives
\begin{equation}
\label{dasym} 
\eqalign{
\bx _{(112)} - \bx _{(12)}&=\frac{A _{(2)}}{H} (\bx _{(12)} - \bx _{(2)} )
                 +\frac{P _{(2)} B _{(1)}}{H}(\bx _{(12)} - \bx _{(1)} )\\
\bx _{(221)} - \bx _{(12)}&=\frac{B _{(1)}}{H} (\bx _{(12)} - \bx _{(1)} )
                 +\frac{Q_{(1)} A _{(2)}}{H}(\bx _{(12)}- \bx _{(2)} ).
}
\end{equation}
The compatibility condition $\bx _{(1122)}=\bx _{(2211)}$ implies that
the functions $A$, $B$, $P$, $Q$ are constrained to \cite{NieszporskiA} 
\begin{equation}
\label{eq:disc-comp-asympt-H}
\frac{A _{(22)}}{A H _{(2)}}=\frac{B _{(11)}}{B H _{(1)}},
\end{equation}
\begin{equation} \label{eq:disc-comp-asympt-HPQ}
\eqalign{
\frac{A _{(22)} H}{A _{(2)} H _{(2)}} (1+B-Q)&=
D _{(1)}- Q _{(1)} C _{(2)},\\
\frac{B _{(11)} H}{B _{(1)} H _{(1)}} (1+A-P)&=
C _{(2)}-P _{(2)}D _{(1)},
}
\end{equation}
where the functions $C$, $D$, $H$ are defined as
\begin{equation}
\label{eq:HCD}
\eqalign{
H&:=1-P _{(2)}  Q _{(1)}, \\
C&:= 1+\frac{A_{(2)}}{H}+\frac{B_{(1)} P_{(2)}}{H},\\
D&:= 1+\frac{B_{(1)}}{H}+\frac{A_{(2)} Q_{(1)}}{H}.
}
\end{equation}

Let us introduce \cite{NieszporskiA} the discrete 
canonical tangent fields $\bW$ and $\bZ$ of the asymptotic lattice $\bx$ by
\begin{equation}
\label{dtan0} \eqalign{
\bx _{(12)} - \bx _{(2)} =\alpha \bW  ,\\
\bx _{(12)} - \bx _{(1)} =\beta \bZ,
}
\end{equation}
where functions $\alpha$ and $\beta$ are defined by
\begin{equation}
\label{ab} \eqalign{
\beta _{(2)} = \frac{B _{(1)}}{H} \beta  ,\\
\alpha _{(1)} = \frac{A _{(2)}}{H} \alpha.
}
\end{equation}
Equations \eqref{dasym} take the form
\begin{equation}
\label{dtan1}
\eqalign{
\Delta_1 \bW &= {\mathcal P} \bZ ,\\
\Delta_2 \bZ &= {\mathcal Q} \bW,
}
\end{equation}
in terms of fields $\bW$ and $\bZ$, where
\begin{equation}
\label{dpq} \eqalign{
{\mathcal P}= \frac{P _{(2)} B _{(1)}}{A _{(2)}} \frac{\beta}{\alpha},
\\
{\mathcal Q}= \frac{Q _{(1)} A _{(2)}}{B _{(1)}} \frac{\alpha}{\beta};
}
\end{equation}
notice that $H=1-{\mathcal P}{\mathcal Q}$.
\begin{Rem}
The first order system \eqref{dtan1} appears, for example, as the
linear problem of the two dimensional quadrilateral 
lattice \cite{DoSaMQL} (see also \ref{app:QL}). 
We will use this fact in Section \ref{sec:FR} where we define the discrete
analogue of the isothermally-asymptotic (Fubini--Ragazzi) nets.
\end{Rem}

\subsection{The discrete Moutard equation and the
Lelieuvre representation of the asymptotic lattices}
It can be shown \cite{KoPi,NieszporskiA} that a suitable rescaled 
normal field $\bN$ to generic asymptotic lattice $\bx$ (for more detailed 
discussion see \cite{NieszporskiA}) is
connected with the lattice itself by the discrete analogue of the Lelieuvre
formulas
\begin{equation}
\label{eq:D-L}
\eqalign{
\Delta_1\bx &= \bN_{(1)}\times\bN, \\
\Delta_2\bx &= \bN\times\bN_{(2)}.
}
\end{equation}
Moreover, there exists a function $F$ such that the normal vector field $\bN$
satisfies the discrete analogue of the
Moutard equation \cite{Nimmo,Schief-DAS}
\begin{equation} \label{eq:Moutard-d}
\bN_{(12)} + \bN = F (\bN_{(1)} + \bN_{(2)}) .
\end{equation}
\begin{Rem}
\label{RDS}
There is an alternative version of the Lelieuvre type representation of  
asymptotic lattices and of the Moutard equation which differs from
\eqref{eq:D-L}-\eqref{eq:Moutard-d} only by a change of signs
\begin{equation}
\label{DME11} 
\bN_{(12)}-\bN=F (\bN_{(1)}-\bN_{(2)}).
\end{equation}
\begin{equation} 
\label{DLF1}\eqalign{
   \Delta_1 \bx= \bN_{(1)} \times \bN ,\\
   \Delta_2  \bx= \bN_{(2)} \times \bN, 
}
\end{equation}
see \cite{NieszporskiA} for details. This minor modification becomes
important when discussing generation of additional dimensions of the lattice
by the Darboux-type transformations. 
\end{Rem}
Notice that, due to the Lelieuvre formulas \eqref{eq:D-L},
there exist functions $\gamma$ and $\delta$ such that
the normal $\bN$ satisfies the linear system
\begin{equation}
\label{DME1N}
\eqalign{
\bN _{(11)} - \bN _{(1)}&=A (\bN _{(1)} - \bN )-P (\bN _{(12)} - 
\bN _{(1)})+\gamma \bN_{(1)}, \\
\bN _{(22)} - \bN _{(2)}&=B (\bN _{(2)} - \bN )-Q (\bN _{(12)} - 
\bN _{(2)})+\delta \bN_{(2)}.
}
\end{equation}
The compatibility of equations \eqref{eq:Moutard-d}-\eqref{DME1N}
gives the relations between the functions $F,\gamma,\delta$ with the fields
$A,B,P,Q$ of the asymptotic lattice $\bx$
\begin{equation} \label{FF}
F F _{(1)}=\frac{A _{(2)}}{A H}, \qquad
F F _{(2)}=\frac{B _{(1)}}{B H},
\end{equation}
\begin{equation} \label{eq:gd}
\eqalign{
\gamma&=-1-A-P+\frac{C}{ F _{(1)}}, \\
\delta&=-1-B-Q+\frac{D}{ F _{(2)}}.
} 
\end{equation}
Notice that the compatibility condition of equations \eqref{FF} is provided
by equation \eqref{eq:disc-comp-asympt-H}.

\subsection{The discrete Moutard transformation}
Given \cite{Nimmo,Schief-DAS}
a scalar solution $\Theta$ of the Moutard  equation 
\eqref{eq:Moutard-d}
\begin{equation} \label{eq:Moutard-d-Theta}
\Theta_{(12)} + \Theta = F (\Theta_{(1)} + \Theta_{(2)}) ,
\end{equation}
then the solution $\bN'$ of the system of equations
\begin{equation}
\label{eq:dmt}
\eqalign{
(\bN'_{(1)}\mp\bN )=\frac{\Theta}{\Theta_{(1)}}(\bN' \mp \bN_{(1)}) \\
(\bN'_{(2)}\pm \bN )=\frac{\Theta}{\Theta_{(2)}}(\bN' \pm \bN_{(2)})
}
\end{equation}
satisfies equation \eqref{eq:Moutard-d} with the transformed
potential
\begin{equation} \label{eq:dmt-F}
F'=\frac{\Theta_{(1)}\Theta_{(2)}}{\Theta\;\Theta_{(12)}}F.
\end{equation}
\begin{Rem} \label{rem:M3}
We consider \cite{DoliwaA,NieszporskiA} two possibilities of signs in the 
Moutard
transformation in order
(i) to preserve the symmetry between the variables 
$m_1$ and $m_2$,  
(ii) to interprete the transformation direction (denoted by prime) 
as a shift in the third variable (see Remark\ref{RDS}), and 
(iii) to reproduce the discrete Moutard equation 
in the superposition formula.
\end{Rem}

The algebraic superposition formula for two Moutard transformations is given
in the following result \cite{DoliwaA,NieszporskiA}
\begin{Th} \label{th:sup-M}
Let $\bN^{(1)}$ be the upper-sign Moutard transform of $\bN$ with respect
to $\Theta^1$, $\bN^{(2)}$ be the lower-sign Moutard transform of $\bN$ 
with respect to $\Theta^2$ and  $\Xi$ be the one parameter family of
solutions of the system
\begin{equation}
\label{eq:Xi}
\eqalign{
\Delta_1\Xi&=\Theta^1_{(1)}\Theta^2 - \Theta^2_{(1)}\Theta^1,\\
\Delta_2\Xi&=\Theta^2_{(2)}\Theta^1 - \Theta^1_{(2)}\Theta^2.
}
\end{equation}
Then the function $\bN^{(12)}$, given by
\begin{equation} \label{eq:superp-M-d}
\bN^{(12)} + \bN = \frac{\Theta^1 \Theta^2}{\Xi}(\bN^{(1)}+\bN^{(2)}),
\end{equation}
is simultaneously the lower-sign Moutard transform of $\bN^{(1)}$ with 
respect to $\Theta^{2(1)}=\Xi / \Theta^1$ and the upper-sign 
Moutard transform of $\bN^{(2)}$ with respect to 
$\Theta^{1(2)}=\Xi / \Theta^2$.
\end{Th}
\begin{Rem} 
When interpreting the
transformation shifts (upper indices in brackets) as shifts in discrete
variables the formula \eqref{eq:superp-M-d} is of the form of the
discrete Moutard equation. 
\end{Rem}

\subsection{The discrete $W$-congruences}
It can be checked directly \cite{NieszporskiA,DoliwaA} that the lattice
$\bx'$ (we still use the $\pm$ convention of the
Moutard transformation) defined by the formula
\begin{equation} \label{eq:x'}
\bx'=\bx \pm \bN'\times\bN,
\end{equation}
is a new asymptotic lattice with the normal $\bN'$ entering in its
Lelieuvre representation. 
\begin{Rem}
Notice again the correspondence between the shifts generated by the Moutard
transformation and the shifts in the discrete variables $m_i$, $i=1,2$.
Namely, in the notation of Theorem \ref{th:sup-M}, the transformation
formulas 
\begin{equation}
\eqalign{
\bx^{(1)}-\bx&= \bN^{(1)}\times\bN,\\
\bx^{(2)}-\bx&= \bN\times\bN^{(2)},
}
\end{equation}
are of the form of the Lelieuvre representation. 
\end{Rem}

The translation of $\bx'$ by a constant vector still is an
asymptotic lattice with the normal $\bN'$. However the lattice $\bx'$
defined in \eqref{eq:x'} helps to define a certain family of lines called 
the discrete $W$ (from Weingarten) congruences \cite{NieszporskiA,DoliwaA}.
The family of straight lines connecting 
$\bx$ and $\bx'$ is tangent to
both asymptotic lattices and has 
analogous  properties to those of the $W$  
congruences known in the theory
of transformations of  asymptotic nets \cite{Bianchi}. 

\begin{Def}[\cite{DoliwaA}] \label{def:DW-cong}
By a {\it discrete W--congruence} we mean a two-parameter family of straight
lines connecting two asymptotic lattices in such a way that the lines
are tangent to both lattices in the corresponding points.
\end{Def}
\begin{Rem}
It can be shown \cite{DoliwaA} that any discrete W--congruence can be 
constructed via the discrete Moutard transformation.
\end{Rem}
The permutability property of superpositions of the Moutard transformation
implies the corresponding permutability of the transformations of the
asymptotic lattices. 
The asymptotic lattice $\bx^{(12)}$, corresponding to the superposition of
the two Moutard transformations in Theorem \ref{th:sup-M}, is given by
\begin{equation}
\bx^{(12)}=\bx+\frac{\Theta^1 \Theta^2}{\Xi}\bN^{(1)}\times\bN^{(2)}.
\end{equation}

\subsection{Discrete Jonas formulas\label{sec:Jonas}}
In this section we present \cite{NieszporskiA}
another useful description, introduced by Jonas \cite{JonasR} in the
continuous case, of the transformation
of asymptotic lattices.

Let $\bN'$ be a transform of  $\bN$ under the discrete Moutard 
transformation (we consider here the upper-sign transformation only).
Define $x^a$, $a=1,2,3$ as coefficients of the decomposition
of  $\Theta\bN'$ in the basis 
$\{ \bN, \bN _{(1)}, \bN _{(2)}\}$ 
\begin{equation} \label{eq:Jonas}
\Theta \bN' = x^1 \bN _{(1)} + x^2 \bN _{(2)} + x^3 \bN.  
\end{equation} 
After substitution of the above expression
into the discrete Moutard transformation we obtain that the
coefficients $x^a$ satisfy six equations, which can be splitted into two parts: 
the following linear system for $x^1$ and $x^2$
\begin{equation}
\label{eq:linJ}
\eqalign{
x^1 _{(2)}-Q x^2 _{(2)}&= \frac{1}{F} x^1, \\
\label{eq:J2}
x^2 _{(1)}-P x^1 _{(1)}&= \frac{1}{F} x^2,
}
\end{equation}
and the remaining equations
\begin{equation} \label{eq:Jx3t}
\eqalign{
x^3 + \Theta _{(1)} & =-A x^1 _{(1)}-\frac{1}{F} x^2,\\
x^3 - \Theta _{(2)} & =-B x^2 _{(2)}-\frac{1}{F} x^1,\\
x^3 _{(1)} +\Theta & =  -(\gamma+1+A+P) x^1 _{(1)} + x^1-x^2, \\
x^3 _{(2)} -\Theta & = - (\delta+1+B+Q)x^2_{(2)}+ x^2-x^1 .
}
\end{equation}
The new normal $\bN'$ satisfies the primed analogue of equations
\eqref{eq:Moutard-d} and \eqref{DME1N}, 
and the primed functions are related to the non-primed ones via
\begin{equation} \label{eq:ABPQ-J}
\eqalign{
P'&=(-P+\frac{S}{L}\frac{x^2}{F})\frac{\Theta _{(12)}}{\Theta _{(11)}},\\
Q'&=(-Q+\frac{T}{L}\frac{x^1}{F})\frac{\Theta _{(12)}}{\Theta _{(22)}},\\
A'&=A(1+\frac{S}{L} x^1 _{(1)}) \frac{\Theta }{\Theta _{(11)}},\\
B'&=B(1+\frac{T}{L} x^2 _{(2)}) \frac{\Theta }{\Theta _{(22)}},
}
\end{equation}
in which
\begin{equation}
\label{DdefJ}
\eqalign{
S&:=\Theta _{(11)} - (\gamma+1+A+P) 
\Theta _{(1)}+ A \Theta + P \Theta _{(12)},\\
T&:=\Theta _{(22)} - (\delta+1+B+Q)\Theta _{(2)}+ 
B \Theta + Q \Theta _{(12)},\\
L&:=x^1 \Theta _{(1)} +x^2 \Theta _{(2)}+x^3 \Theta.
}
\end{equation}
The Jonas formulation gives an alternative way to construct transformations
of asymptotic lattices.
\begin{Th}
Consider an asymptotic lattice $\bx$ and its normal $\bN$ connected by the
Lelieuvre representation. Any non-zero solution $(x^1,x^2)$ of the 
linear system \eqref{eq:linJ} leads, via equations \eqref{eq:Jx3t},
to functions $x^3$ and $\Theta$ such that:\\
i) $\Theta$ satisfies the Moutard equation of $\bN$;\\
ii) $\bN'$, given by \eqref{eq:Jonas} and $\bx'$, given by the upper-sign
version of \eqref{eq:x'}, are the corresponding transforms of $\bN$ 
and $\bx$. 
\end{Th}

\subsection{The Pl\"{u}cker geometry approach to asymptotic lattices and $W$
congruences}
In this section we present \cite{DoliwaA} the theory of asymptotic lattices 
and their transformations 
in the language of the line geometry of Pl\"ucker (see \ref{app:lg}).

Denote by $\gp_i$, $i=1,2$ the bi-vectors representing the asymptotic lines
of the lattice $\bx$, i.e., the lines passing
through points $\bx$ and $\bx_{(i)}$
\begin{equation}
\gp_i = \left( \begin{array}{l} \bx \\ 1 \end{array} \right) \wedge
\left( \begin{array}{l}  \bx_{(i)} \\ 1  \end{array} \right) , \qquad i=1,2.
\end{equation} 
Equations \eqref{eq:da1} imply the following linear system
\begin{equation}
\eqalign{
\label{eq:p}
\gp_{1(1)} & = A \gp_1 + P\gp_{2(1)},\\
\gp_{2(2)} & = B \gp_2 + Q\gp_{1(2)}.
}
\end{equation}

\begin{figure}
\begin{center}
\epsffile{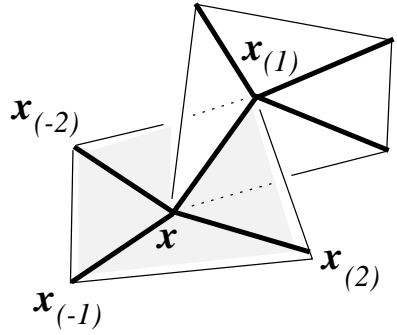}
\end{center}
\caption{Asymptotic directions as focal lattices of the 
isotropic congruence}
\label{fig:izot-cong}
\end{figure}

The planar pencils of straight lines are represented in the Pl\"{u}cker
geometry by isotropic (i.e., contained in $\cQ_P$) lines. Therefore the
tangent planes of the asymptotic lattice are represented by two-parameter
family of isotropic lines. Since two neighbouring tangent planes, 
in $\bx$ and $\bx_{(i)}$, intersect along the asymptotic line represented by
$\gp_i$, then the corresponding two isotropic lines have one point in common
(see Figure \ref{fig:izot-cong}).
Using the terminology of the theory of quadrilateral lattices (see \ref{app:QL})
the above considerations can be summarized as follows.
\begin{Th}[\cite{DoliwaA}] \label{th:D-A}
A discrete asymptotic net in $\PP^3$, viewed as the envelope of its tangent 
planes, corresponds to a congruence of isotropic lines
of the Pl\"{u}cker quadric 
$\cQ_P$. The focal lattices of the congruence represent asymptotic 
directions of the lattice.
\end{Th}
\begin{Rem}
The lattices in $\cQ_P$, given by the bi-vectors $\gp_1$ and $\gp_2$, 
which represent two families of asymptotic tangents 
of the asymptotic lattice, are Laplace transforms of each other and satisfy
the following discrete Laplace equations
\begin{equation}
\eqalign{
\gp_{1(12)} &= \frac{P_{(2)} B_{(1)}}{P H}\gp_{1(1)} +
\frac{A_{(2)}}{H}\gp_{1(2)} -\frac{P_{(2)} B_{(1)}A}{P H}\gp_{1},\\
\gp_{2(12)} &= \frac{Q_{(1)} A_{(2)}}{Q H}\gp_{2(2)} +
\frac{B_{(1)}}{H}\gp_{2(1)} -\frac{Q_{(1)} A_{(2)}B}{Q H}\gp_{2}.
}
\end{equation}
\end{Rem}
Finally, let us consider the line-geometric interpretation of the $W$
congruences.
\begin{Rem}
In contrary to the continuous case, the discrete $W$ congruence is not a
congruence in the sense of the definition used in the theory of transformations
of quadrilateral lattices (see discussion in \cite{DoliwaA}). 
\end{Rem}
The bi-vector
\begin{equation}
\gq = \Theta\left( \begin{array}{l} \bx \\ 1 \end{array} \right) \wedge
\left( \begin{array}{l}  \bx' \\ 1  \end{array} \right) , 
\end{equation} 
represents, in a convenient gauge, the line connecting $\bx$ and $\bx'$.
From the decomposition \eqref{eq:Jonas} (we take again
$\bx'$ from formula \eqref{eq:x'} with the upper sign) and the
Lelieuvre representation \eqref{eq:D-L} we obtain that
\begin{equation}
\gq= x^1\gp_1 -x^2\gp_2.
\end{equation}
The linear problems \eqref{eq:linJ} and
\eqref{eq:p} and equations \eqref{FF}
imply that $\gq$ satisfies the Laplace equation 
\begin{equation}
\label{ALPKQ}
\gq_{(12)}= \frac{x^2_{(2)}}{x^2}  F^2 B \gq_{(1)}+
\frac{x^1_{(1)}}{x^1}  F^2 A \gq_{(2)}-
\frac{x^1_{(1)}x^2_{(2)}}{x^1 x^2}
F^2 AB\gq.
\end{equation}
\begin{Th}[\cite{DoliwaA}] Discrete $W$ congruences are represented by two
dimensional quadrilateral lattices in the Pl\"ucker quadric $\cQ_P$.
\end{Th}
\begin{Rem}
The $W$ congruences provide an example of quadrilateral lattices
subjected to a quadratic constraint. A general theory of such quadratic
reductions of quadrilateral lattices was 
studied in \cite{DolQR}.
\end{Rem}
\begin{Rem}
Since the points of intersection of the Pl\"ucker quadric with a plane
represent a regulus (one family of generators of a ruled quadric in 
$\PP^3$) then \cite{DoliwaA} four neighbouring lines of a $W$ congruence are
lines of the same regulus. This property of $W$ congruences can be used to
define them without using the notion of the asymptotic lattice. 
\end{Rem}

\section{Integrable reductions of asymptotic lattices}
\label{sec:R}
In this section we consider two basic integrable reductions of the
asymptotic lattices: the Bianchi lattice and the Fubini-Ragazzi lattice,
which are the integrable discretizations of the Bianchi and Fubini-Ragazzi
surfaces respectively. These integrable reductions are obtained: \\
 i) imposing suitable nonlinear constraints on the geometric data of the
asymptotic lattice (hence obtaining a nonlinear system of equations,
characterized by the discrete Moutard equation and by the nonlinear
constraint); and then\\
 ii) showing that these constraints are preserved by the discrete Moutard 
transformation (which therefore allows one to obtain, through a
sequence of linear steps only, a new solution of the above nonlinear
lattice equations from a given one).

\subsection{The discrete analogue of the Bianchi--Ernst system}
Let us equip $\RR^3$ with the scalar product "$\cdot$" 
\begin{equation}
{\bi{A}}\cdot{\bi{B}}:=A_0B_0 + \epsilon (A_1B_1 + A_2
B_2), \qquad \epsilon = \pm 1.
\end{equation} 
In this section we discuss (see \cite{NDS} for more details)
the integrability of the Bianchi--Ernst reduction 
\begin{equation} \label{eq:BE-d-constr}
(\bN_{(12)} + \bN)\cdot(\bN_{(1)} + \bN_{(2)}) =U(m_1)+V(m_2),
\end{equation}
of the Moutard equation \eqref{eq:Moutard-d}; here $U(m_1)$ is a function of
$m_1$ only and $V(m_2)$ is a
function of $m_2$. 
\begin{Rem}
For simplicity of considerations we assume that $U(m_1) + V(m_2)>0$. 
\end{Rem} 
In order to construct a suitable reduction of the Moutard transformation
which would preserve the constraint \eqref{eq:BE-d-constr} it is important
to notice the following condition. 
\begin{Th}
If $\bN$ and $\bN'$ are connected by the discrete Moutard transformation
\eqref{eq:dmt}, then the condition
\begin{equation}
\label{eq:B-N-N'}
 (\bN_{(12)} + \bN )\cdot (\bN_{(1)} + \bN_{(2)}) =
 (\bN'_{(12)} + \bN')\cdot(\bN'_{(1)} + \bN'_{(2)})
\end{equation}
is equivalent to the constraint \eqref{eq:BE-d-constr} supplemented by
equations
\begin{equation}
\label{eq:B-N-N'12} \eqalign{
(\bN' _{(1)} \mp \bN)\cdot (\bN' \mp \bN _{(1)})&=U(m_1)\mp k,\\
(\bN' _{(2)} \pm \bN)\cdot (\bN' \pm \bN _{(2)})&=V(m_2) \pm k,
}
\end{equation}
where $U(m_1)$ and $V(m_2)$ are functions of single variables only 
and $k$ is a constant.
\end{Th}
\begin{Rem} \label{rem:MB3}
Notice that, if we consider the transformation direction (denoted by prime) 
as a shift in a third variable and make use of the freedom of the form of
the Moutard equation (see Remark\ref{RDS}), then the constraint
\eqref{eq:B-N-N'12} on the discrete Moutard 
transformation is itself of the form of discrete Bianchi-Ernst constraint
\eqref{eq:BE-d-constr}.
\end{Rem}
The integrability of the Bianchi--Ernst lattices is the consequence of the
following result \cite{NDS}. 
\begin{Th} \label{th:dBtr}
Given a solution $\bN$ of the Bianchi--Ernst system 
\eqref{eq:Moutard-d}, \eqref{eq:BE-d-constr} and given
the $\epsilon$-unit vectors $\bn_1$, $\bn_2$ (i.e., $\bn_1\cdot
\bn_1=\bn_2\cdot\bn_2=\epsilon$) orthogonal to $\bN_{(12)}+\bN=:\bn_0$ and 
to each other, then:\\
i) The linear system 
\begin{equation} \label{eq:lin-BE}
\eqalign{
\fl \left( \begin{array}{c} 
\Theta \\ \Theta_{(1)} \\ \Theta_{(2)} \\ y^1 \\ y^2
\end{array}\right)_{(1)}=
\left( \begin{array}{ccccc}
0&1&0&0&0 \\
\frac{Y_{(1)}\frac{p_0^0}{F}-b}{a_{(1)}}&
\frac{b F- Y_{(1)}(\frac{Y+b}{Y}p_0^0-\frac{1}{F _{(1)}})}{a_{(1)}}&
\frac{b}{a_{(1)}}(F-\frac{Y_{(1)}}{Y}p_0^0)&
\mp \frac{Y_{(1)}}{a_{(1)}}p_{1}^0&\mp\frac{Y_{(1)}}{a_{(1)}}p_{2}^0 \\
-1&F&F&0&0\\
\mp \frac{p_0^{1}}{F} & \pm\frac{Y+b}{Y}p_0^{1} &
\pm\frac{b}{Y}p_0^{1} & p_{1}^{1} & p_{2}^{1}\\
\mp\frac{p_0^{2}}{F} & \pm\frac{Y+b}{Y}p_0^{2} &
\pm\frac{b}{Y}p_0^{2} & p_{1}^{2} & p_{2}^{2}
\end{array}\right)
\left( \begin{array}{c} 
\Theta \\ \Theta_{(1)} \\ \Theta_{(2)} \\ y^1 \\ y^2
\end{array}\right),\\ \\
\fl \left( \begin{array}{c}
\Theta \\ \Theta_{(1)} \\ \Theta_{(2)} \\ y^1 \\ y^2
\end{array}\right)_{(2)}=
\left( \begin{array}{ccccc}
0&0&1&0&0\\
-1&F&F&0&0\\
\frac{Y_{(2)}\frac{q_0^0}{F}-a}{b_{(2)}}&
\frac{a}{b_{(2)}}(F-\frac{Y_{(2)}}{Y}q_0^0)&
\frac{a F- Y_{(2)}(\frac{Y+a}{Y}q_0^0-\frac{1}{F _{(2)}})}{b_{(2)}}&
\pm\frac{Y_{(2)}}{b_{(2)}}q_{1}^0&\pm\frac{Y_{(2)}}{b_{(2)}}q_{2}^0
\\
\pm\frac{q_0^{1}}{F}&\mp\frac{a}{Y}q_0^{1}&\mp\frac{a+Y}{Y}q_0^{1}
&q_{1}^{1}&q_{2}^{1}\\
\pm\frac{q_0^{2}}{F}&\mp\frac{a}{Y}q_0^{2}&\mp\frac{a+Y}{Y}q_0^{2}
&q_{1}^{2}&q_{2}^{2}
\end{array}\right)
\left( \begin{array}{c}
\Theta \\ \Theta_{(1)} \\ \Theta_{(2)} \\ y^1 \\ y^2
\end{array}\right),
}
\end{equation}
where $F$, $Y$, $a$, $b$ are given by equations 
\begin{equation} \label{eq:F-red}
F=\frac{U(m_1) + V(m_2)}{(\bN_{(1)} + \bN_{(2)})\cdot
(\bN_{(1)} + \bN_{(2)})} ,
\end{equation}
\begin{equation} \label{eq:Yab-red}
Y=U(m_1)+V(m_2), \qquad a=U(m_1) \mp k, \qquad b=V(m_2) \pm k.
\end{equation}
and $p^A_B$, $q^A_B$, $A,B=0,1,2$ are defined by the 
unique decompositions (we use the summation convention)
\begin{equation} \label{eq:p-q}
\bn_A=p_A^B \bn_{B(1)}, \qquad \bn_A=q_A^B \bn_{B(2)},
\end{equation}
is compatible.\\
ii) The solution $(\Theta,\Theta_{(1)},\Theta_{(2)},y^1,y^2)$ of the 
system \eqref{eq:lin-BE} satisfies the constraint
\begin{equation} \label{eq:constraint}
\fl \epsilon[(y^1)^2+(y^2)^2]+\frac{Y}{F}\Theta^2 +
FY\left( -\frac{a}{Y}\Theta_{(1)} +
\frac{b}{Y}\Theta_{(2)}\right)^2 -
2\Theta\left( a\Theta_{(1)}+b\Theta_{(2)} \right) =0,
\end{equation}
provided that such a constraint is satisfied at 
the initial point.\\  
iii) Given the solution $(\Theta,\Theta_{(1)},\Theta_{(2)},y^1,y^2)$ of the 
system \eqref{eq:lin-BE} satisfying the constraint
\eqref{eq:constraint}, then $\bN'$, constructed via equation 
\begin{equation}
\label{eq:xA}
\bN'= \frac{1}{2}\left(\pm \bN_{(1)} \mp \bN_{(2)} \right) +
\frac{y^A}{2\Theta}\bn_A,
\end{equation} 
with $y^0$ given by 
\begin{equation} \label{eq:x0}
y^0 = \frac{\mp\Theta_{(1)} (U \mp k) \pm \Theta_{(2)}  (V \pm k)}{U+V},
\end{equation} 
is a new solution of the discrete
Bianchi--Ernst system.
\end{Th}
\begin{Rem}
The parameter $k$, present in the linear system
\eqref{eq:lin-BE}, is called the transformation parameter.
The linear system can also be interpreted
as a nonstandard Lax pair (zero curvature representation) of the discrete
Bianchi-Ernst system, with spectral parameter $k$. 
\end{Rem}

In our recent paper we announced the theorem on the permutability of the
superposition of discrete Bianchi transformations.
\begin{Th}[\cite{NDS}] \label{th:superposition}
Given a solution $\bN$ of the Bianchi--Ernst system and given
two transforms of it: 
the upper-sign transform $\bN^{(1)}$, with the transformation 
parameter $k^1$, and  the lower-sign transform
$\bN^{(2)}$, with the transformation parameter $k^2$. Then there exists the
unique solution $\bN^{(12)}$ of the Bianchi--Ernst system, given in 
algebraic terms by
\begin{equation} \label{eq:BE-superp}
\bN^{(12)} = -\bN +
\frac{k^1+k^2}{(\bN^{(1)}+\bN^{(2)})\cdot(\bN^{(1)}+\bN^{(2)})}
(\bN^{(1)}+\bN^{(2)}),
\end{equation}
which is simultaneously the lower-sign transform of $\bN^{(1)}$, with 
the transformation parameter $k^2$, and the upper-sign transform of 
$\bN^{(2)}$, with the transformation parameter $k^1$.
\end{Th}
\begin{proof}
We first show that $\bN^{(12)}$ defined in \eqref{eq:BE-superp}
is a Moutard transform of $\bN^{(1)}$ and of $\bN^{(2)}$. To do that we have
to check that the function $\Xi$, which due to 
equation \eqref{eq:superp-M-d} must be of the form
\begin{equation}
\label{Xi-B}
\Xi= \frac{\Theta^1  \Theta^2}{k^1+k^2}
(\bN^{(1)} + \bN^{(2)})\cdot(\bN^{(1)} + \bN^{(2)}) ,
\end{equation}
does satisfy equations \eqref{eq:Xi}. This can be verified directly using
formulas \eqref{eq:dmt} and \eqref{eq:B-N-N'12} applied to $\bN^{(1)}$ and 
$\bN^{(2)}$.

Now, since $\bN^{(12)}$ exists and describes an asymptotic lattice,
it is enough to show that the equations
\eqref{eq:B-N-N'12} with the correct constant and sign
apply also for the pair 
($\bN^{(12)}$, $\bN^{(1)}$) and for the pair ($\bN^{(12)}$, $\bN^{(2)}$), 
what can be done by direct calculation using \eqref{eq:dmt}, 
\eqref{eq:BE-d-constr} and \eqref{eq:BE-superp}.
\end{proof}
\begin{Rem}
Notice that the superposition formula \eqref{eq:BE-superp} for
the Bianchi--Ernst system reproduces the 
Bianchi--Ernst system itself, after replacing the upper transformation
indices by the lower translation ones. 
\end{Rem}
\begin{Rem}
If we treat the transformations as shifts in additional parameters (denoted by
$m^1$ and $m^2$), then
the vector function 
$\bN(m_1,m_2,m^1,m^2)$ satisfies the discrete Bianchi--Ernst system in 
every pair of parameters (see also Remarks \ref{rem:M3} and 
\ref{rem:MB3}). Similar consideration in
connection with the relation between the superposition formula for the
discrete Tzitzeica equation and the self-dual Einstein spaces appeared 
in \cite{Schief-DAS}.
\end{Rem}

\begin{Rem}
On introducing vector field ${\bnu}:= \frac{\bN_{(12)}+\bN}{\sqrt{F}}$ 
it is easy to show that ${\bnu}$ is a solution of
\begin{equation}\label{eq:n0form2}
\frac{{ \bnu}_{(12)}}{\sqrt{F_{(12)}}} + \frac{{ \bnu}}
{\sqrt{F}} = 
\sqrt{F_{(1)}} { \bnu}_{(1)} + \sqrt{F_{(2)}} 
{ \bnu}_{(2)},
\end{equation}
\begin{equation} \label{eq:NN-Bd}
{ \bnu}\cdot{ \bnu} =U(m_1)+V(m_2)=:r;
\end{equation}
provided that $\bN$ is a solution of
\eqref{eq:BE-d-constr}-\eqref{eq:Moutard-d}.

\end{Rem}

\subsection{The discrete analogue of the Fubini--Ragazzi system}
\label{sec:FR}
Let us impose the symmetric reduction condition \eqref{eq:symm-QQ}
on the equation of the discrete tangent canonical fields \eqref{dtan1}
of a  asymptotic lattice, i.e., 
\begin{equation} \label{dFRc}
\frac{\rho_{(12)} \rho}{\rho_{(1)}\rho_{(2)}} =\frac{H_{(2)}}{H_{(1)}} , \qquad
\rho:= \frac{{\mathcal P}}{{\mathcal Q}}.
\end{equation}
The constraint \eqref{dFRc} with equations
\eqref{eq:disc-comp-asympt-H} and \eqref{eq:disc-comp-asympt-HPQ}
gives the discrete version of the  Fubini-Ragazzi system 
\cite{Fubini,NieszporskiA}.
The Darboux-B\"acklund transformation for the 
discrete  Fubini-Ragazzi comes directly from the discrete Jonas formulas
of Section \ref{sec:Jonas}
(details can be found in \cite{NieszporskiD}).
\begin{Th} Consider a Fubini--Ragazzi lattice $\bx$ with the normal $\bN$,
and let the function $\zeta$ be solution of the system 
\begin{equation}
\label{paramk}
\eqalign{
\frac{\zeta_{(2)}}{\zeta}=\frac{H (QF^2)_{(12)}}{Q_{(1)}},\\
\frac{\zeta_{(1)}}{\zeta}=\frac{H (PF^2)_{(12)}}{P_{(2)}},
}
\end{equation}
(i.e., $\zeta$ is given up to a constant parameter, say $k$) 
then:\\
i) The linear system
\small{
\begin{equation}
\label{yi}
\fl
\begin{array}{l}
\left( 
\matrix{x^1 \cr x^2 \cr x^3 \cr \Theta_{(1)} 
\cr \Theta_{(2)} \cr \Theta} 
\right)_{(1)} =
\left( \matrix{ 0 & -\frac{1}{FA}  & -\frac{1}{A}    & -\frac{1}{A}  &  0  
		&  0   \cr
                0 & \frac{A-P}{FA} & -\frac{P}{A}    & -\frac{P}{A}  &  0  
		&  0   \cr
           1 & \frac{C}{AFF_{(1)}}-1 & \frac{C}{AF_{(1)}}& 
	   \frac{C}{AF_{(1)}} & 0 & -1  \cr
           0 &   \frac{-\zeta}{FAQ_{(1)}(F_{(1)})^2} & 
	   \frac{-\zeta}{A Q_{(1)} (F_{(1)})^2}&
\frac{C-PFF_{(1)}}{F_{(1)}} -\frac{\zeta}{A Q_{(1)} (F_{(1)})^2}
 &-PF&P-A\cr
                0&0&0&F&F&-1\cr
                0&0&0&1&0&0\cr} \right)
\left( \matrix{x^1 \cr x^2 \cr x^3 \cr \Theta_{(1)} 
\cr \Theta_{(2)} \cr \Theta} \right) 
\\
\\
\left(
\matrix{x^1 \cr x^2 \cr x^3 \cr \Theta_{(1)} 
\cr \Theta_{(2)} \cr \Theta}
\right)_{(2)} =
\left( \matrix{ \frac{B-Q}{FB}   & 0 & -\frac{Q}{B} & 0 &  \frac{Q}{B} &   0    \cr
                -\frac{1}{FB}   &  0 & -\frac{1}{B} & 0 &  \frac{1}{B} &   0    \cr
          \frac{D}{BFF_{(2)}}-1 & 1 & \frac{D}{BF_{(2)}} & 0 &
	  -\frac{D}{BF_{(2)}} & 1 \cr
                0&0&0&F&F&-1\cr
  \frac{-\zeta}{FBP_{(2)}(F_{(2)})^2}   &    0  & \frac{-\zeta}{BP_{(2)}(F_{(2)})^2} &-QF& 
\frac{D-QFF_{(2)}}{F_{(2)}}+  \frac{\zeta}{BP_{(2)}(F_{(2)})^2} 
&  Q-B   \cr
                0&0&0&0&1&0\cr} \right) 
\left( \matrix{x^1 \cr x^2 \cr x^3 \cr \Theta_{(1)} 
\cr \Theta_{(2)} \cr \Theta} \right) ,
 \end{array}
\end{equation}
}
is compatible.\\
ii) Given the solution 
$(x^1, x^2, x^3, \Theta_{(1)}, \Theta_{(2)} , \Theta)$ of the 
system \eqref{yi} then 
$\bx'$, constructed via the upper-sign formula \eqref{eq:x'}, with $\bN'$ 
constructed via equation 
\eqref{eq:Jonas}, is a new Fubini--Ragazzi lattice. The corresponding new
solution of equations \eqref{eq:disc-comp-asympt-H},
\eqref{eq:disc-comp-asympt-HPQ}, subjected to the
constraint \eqref{dFRc}, is given by formulas 
\eqref{eq:ABPQ-J}-\eqref{DdefJ}.
\end{Th}
\begin{Rem}
The parameter $k$, present in the linear system
\eqref{yi} via equations \eqref{paramk}, is called the transformation 
parameter.
The linear system can also be interpreted
as the Lax pair (zero curvature representation) of the discrete
Fubini--Ragazzi system, with spectral parameter $k$. 
\end{Rem}
\begin{Rem}
The discrete  analogue of (indefinite) affine spheres 
\cite{BoSchA,BoSchI}, which is described by the discrete
analogue of the Tzitzeica equation
is the particular reduction of the discrete Fubini--Ragazzi lattice
corresponding to
\begin{equation} \eqalign{
C=F_{(1)}(1+A+P),\\
D=F_{(2)}(1+B+Q),
}
\end{equation}
i.e., $\gamma=\delta=0$ in equations \eqref{DME1N}.
\end{Rem}

\appendix

\section{Quadrilateral lattices}
\label{app:QL}

We will need few basic facts from the theory of quadrilateral lattices,
which are the discrete integrable analogues of conjugate nets
\cite{Sauer,DCN,DoSaMQL}. The $N$ dimensional quadrilateral lattice in
$\PP^M$, $2\leq N \leq M$, is
geometrically characterized by the planarity of the elementary quadrilaterals of 
the lattice. In terms of the homogeneous representation 
$\by:\ZZ^N\to\RR^{M+1}_*$ of the lattice, this geometric characterization 
can be algebraically
expressed as a linear constraint between $\by$, $\by_{(i)}$, $\by_{(j)}$ and
$\by_{(ij)}$, where $i\ne j$ and $i,j=1,\dots,N$. In the generic case, 
such a linear relation can be
put in the form of the so called discrete Laplace equation
\cite{DCN,DoSaMQL}
\begin{equation} \label{eq:dL}
\by_{(ij)}=\calA_{ij}\by_{(i)} + \calA_{ji}\by_{(j)} + \calC_{ij}\by, \quad
i\ne j, \quad \calC_{ij}=\calC_{ji}.
\end{equation}

From the theory of the Darboux-type transformations of the quadrilateral
lattices \cite{DoSaMa} we recall that
a $\ZZ^N$-parameter family of lines such that any two neighbouring lines 
intersect is called an ($N$ dimensional) discrete congruence. 
The intersection points of lines of the congruence with their $i$-direction
neighbours define the $i$th focal lattice of the congruence. 
Such focal lattices are, in general, quadrilateral lattices.
Any two focal lattices of the
congruence are connected by the so called Laplace transformation \cite{DCN}.

In the affine gauge the system of Laplace equations \eqref{eq:dL} can be
replaced by the following linear system (see also \cite{BoKo})
\begin{equation}
\Delta_j\bX_i = Q_{ij(j)}\bX_j, \quad i\ne j.
\end{equation}
An important integrable reduction of the quadrilateral lattice is the so
called symmetric reduction \cite{DoSaS}. Among various characterizations of
the symmetric lattice we will use the following constraint
\begin{equation}\label{eq:symm-QQ}
\frac{r_{ij(ij)} r_{ij}}{r_{ij(i)}r_{ij(j)}} = 
\frac{(1 - Q_{ji(i)} Q_{ij(j)})_{(i)} }{(1 - Q_{ji(i)} Q_{ij(j)})_{(j)} } ,
\qquad i\ne j,
\end{equation}
where
\begin{equation} 
r_{ij} := \frac{Q_{ij(j)}}{Q_{ji(i)}} , \qquad i\ne j .
\end{equation}

\section{The line geometry of Pl\"{u}cker}
\label{app:lg}
In the line geometry the primary elements are
straight lines in $\RR^3$. It is convenient
to consider $\RR^3$ as the affine part of the projective space
$\PP^3$ (by the standard embedding $\bx \mapsto [(\bx,1)^T]$),
and study straight lines in that space. 

The line passing through two points $[\bu]$, $[\bv]$ of $\PP^3$, can be 
represented, up to proportionality factor, by a bi-vector 
\begin{equation} \label{eq:bi-v-def}
\gp=\bu\wedge\bv \in \bigwedge^{2}(\RR^{4}) .
\end{equation} 
The space of straight lines in $\PP^3$ can be therefore
identified with a subset of $\PP\left(\bigwedge^{2}(\RR^{4})\right)
\simeq \PP^5$. The necessary and sufficient condition for a non-zero
bi-vector $\gp$ in order 
to represent a straight line is given by the homogeneous equation
\begin{equation} \label{eq:bi-v-simple}
\gp \wedge \gp = 0 .
\end{equation}
If $\boe_1,\dots\boe_4$ is a basis of $\RR^4$ then the following
bi-vectors 
\begin{equation} \boe_{i_1 i_2}=
\boe_{i_1}\wedge \boe_{i_2}, \qquad 1\leq i_1 < i_2\leq 4,
\end{equation}
form the corresponding basis of $\bigwedge^{2}(\RR^{4})$:
\begin{equation}
\gp = p^{12}\boe_{12} + p^{13}\boe_{13} + 
\dots + p^{34}\boe_{34} \; .
\end{equation}
Equation~\eqref{eq:bi-v-simple} rewritten in the Pl\"ucker coordinates
$p^{ij}$ reads
\begin{equation} \label{eq:Pl-quad}
p^{12}p^{34}-p^{13}p^{24}+p^{14}p^{23} = 0 \; ,
\end{equation}
and defines in $\PP^5$ the so-called Pl\"{u}cker 
quadric $\cQ_P$.

\ack A. D. and P. M. S. would like to thank of the organizers of SIDE IV 
meeting for  the
kind invitation and support during the conference.


\section*{References}

\begin{thebibliography}{99}

\bibitem{Bianchi}
Bianchi L 1923 {\it Lezioni di Geometria Differenziale} (Pisa: Spoerri) 

\bibitem{BoPi-DP}
Bobenko A I and Pinkall U  1996 Discrete surfaces with constant 
negative
{G}aussian curvature and the {H}irota equation, {\it J. Diff. Geom. }
{\bf 43} 527--611


\bibitem{BoSchA}
Bobenko A I and Schief W K 1999
    Affine spheres: Discretization via duality relations
    {\it Exp. Math.} {\bf 8} 261-280
    
\bibitem{BoSchI}
\dash 1999
Discrete indefinite affine spheres
{\it Discrete integrable geometry and physics}
(Oxford: Clarendon Press)

\bibitem{BoKo}
Bogdanov L V and  Konopelchenko B G 1995
Lattice and q-difference Darboux-Zakharov-Manakov systems via 
$\bar{\partial}$-dressing method 
{\it J. Phys.} A {\bf 28} L173--L178

\bibitem{DCN}
Doliwa A 1997 Geometric discretisation of the Toda 
system {\it Phys. Lett.} A {\bf 234} 187--192

\bibitem{DolQR}
\dash  1999 Quadratic reductions of quadrilateral lattices  
{\it J. Geom. Phys.} {\bf 30} 169-186

\bibitem{DoliwaA} 
\dash  2001
Discrete asymptotic nets and W-congruences in Pl\"ucker line geometry 
{\it J. Geom. Phys.} {\bf 39}  9--29

\bibitem{DoSaMQL}
Doliwa A and Santini P M 1997
Multidimensional quadrilateral lattices are integrable
{\it Phys. Lett.} A  {\bf 233} 365--372


\bibitem{DoSaS}
\dash 2000
The symmetric, D-invariant and Egorov reductions of the quadrilateral 
lattice 
{J. Geom. Phys.} {\bf 36} 60-102

\bibitem{DoSaMa}
Doliwa A, Santini P M and  Ma\~{n}as M 2000
Transformations of quadrilateral lattices
{\it J. Math. Phys.} {\bf 41} 944-990

\bibitem{DoManSa}
Doliwa A,  Manakov S V and Santini P M 1998
$\bar{\partial}$--reductions of the multidimensional
quadrilateral lattice: the multidimensional circular lattice
{\it Comm. Math. Phys.} {\bf 196} 1-18

\bibitem{Fubini}
Fubini G 1916
Su una classe di congruenze $W$ di caractere proiettiva
{\it Rend. Lincei} (5)  {\bf 25} 144-148

\bibitem{JonasR}
Jonas H 1920 
\"Uber die Konstruktion der $W$-Kongruenzen zu einem gegebenen
 Brennfl\"achenmantel und \"uber die Transformation der $R$-Fl\"achen
{\it J. Deutsch. Math. Ver.} {\bf 29} 40-74


\bibitem{KoPi}
Konopelchenko B G and Pinkall U 2000  Projective generalizations of
  {L}elieuvre's formula {\it Geometriae Dedicata} {\bf 79}  81--99

\bibitem{KoSchi}
Konopelchenko B G and  Schief W K 1998
Three-dimensional integrable lattices in Euclidean spaces: conjugacy and 
orthogonality 
{\it P. Roy. Soc. Lond.} A  {\bf 454} 3075--3104 

\bibitem{Levi}
Levi D and Sym A 1990
Integrable systems describing surfaces of non-constant curvature
{\it Phys. Lett.} A {\bf 149} 381--387

\bibitem{NieszporskiA}
Nieszporski M
On a discretization of asymptotic nets {\it J.  Geom.  Phys.} 
(to be published) 

\bibitem{NieszporskiD}
\dash Weingarten congruences as a source of integrable
systems  (Ph. D. thesis [in Polish])

\bibitem{NDS}
Nieszporski M, Doliwa A and Santini P M 2001 The integrable discretization
of the Bianchi--Ernst system {\it Preprint} nlin.SI/0104065

\bibitem{Nimmo}
Nimmo J J C and Schief W K 1997
Superposition principles associated with the Moutard transformation: 
an integrable discretization of a 2+1-dimensional sine-Gordon system 
{\it Proc. R. Soc. London} A {\bf 453} 255--279

\bibitem{Ragazzi}
Ragazzi E 1921 Sopra una classe di transformazioni delle superficie
isothermo--assintotiche (asymptotiche) ed il loro teorema di permutabilita
{\it Rend. Palermo} {\bf 45} 200-210

\bibitem{Sauer}
Sauer R 1970  
{\it Differenzengeometrie} (Berlin: 
Springer Verlag)

\bibitem{Schief-DAS}
Schief W K 1999 Self-dual Einstein spaces and a discrete Tzitzeica equation.
A permutability theorem link {\it Symmetries and Integrability of 
Difference Equations} Clarkson P A and F W Nijhoff (eds.)
(Cambridge: University Press) 137--148

\bibitem{Tzitzica}
Tzitzeica G 1910 Sur une nouvelle classe des surfaces {\it C. R. Acad. Sci.
Paris} {\bf 150} 585--598
 




\end{thebibliography}
\end{document}